\documentclass[a4paper]{spie}  
\usepackage[]{graphicx}
\usepackage{amstext} 
\usepackage{amssymb} 
\usepackage{amsmath}

\def\arcsec{\hbox{$^{\prime\prime}$}}
\def\degr{\hbox{$^\circ$}}

\title{Narrow-angle astrometry with PRIMA} 

\author{J.~Sahlmann\supit{a}, D.~S\'egransan\supit{a}, A.~M\'erand\supit{b}, N.~Zimmerman\supit{c}, R.~Abuter\supit{d}, B. Chazelas\supit{a}, F.~Delplancke\supit{d}, T. Henning\supit{c}, A.~Kaminski\supit{e}, R.~K\"ohler\supit{c,e}, R.~Launhardt\supit{c}, M. Mohler\supit{c}, F. Pepe\supit{a}, D. Queloz\supit{a}, A. Quirrenbach\supit{e}, S. Reffert\supit{e}, C.~Schmid\supit{d}, N.~Schuhler\supit{b}, T. Schulze-Hartung\supit{c} 
\skiplinehalf
\supit{a}Observatoire de Gen\`eve, Universit\'e de Gen\`eve, 51 Chemin Des Maillettes, 1290 Versoix, Switzerland; \\
\supit{b}European Southern Observatory, Alonso de C\'ordova 3107, Vitacura-Santiago, Chile;\\
\supit{c}Max-Planck-Institut f\"{u}r Astronomie, K\"{o}nigstuhl 17, D-69117 Heidelberg, Germany;\\
\supit{d}European Southern Observatory, Karl-Schwarzschild-Str. 2, 85748 Garching, Germany; \\
\supit{e}ZAH Landessternwarte, K\"onigstuhl 12, 69117 Heidelberg, Germany
}
\authorinfo{Further author information: (Send correspondence to J.S.)\\J.S.: E-mail: johannes.sahlmann@unige.ch, Telephone: +41 223 79 2404}

\begin{document} 
  \maketitle 

\begin{abstract}
The Extrasolar Planet Search with PRIMA project (ESPRI) aims at characterising and detecting extrasolar planets by measuring the host star's reflex motion using the narrow-angle astrometry capability of the PRIMA facility at the Very Large Telescope Interferometer. A first functional demonstration of the astrometric mode was achieved in early 2011. This marked the start of the astrometric commissioning phase with the purpose of characterising the instrument's performance, which ultimately has to be sufficient for exoplanet detection. We show results obtained from the observation of bright visual binary stars, which serve as test objects to determine the instrument's astrometric precision, its accuracy, and the plate scale. Finally, we report on the current status of the ESPRI project, in view of starting its scientific programme. 
\end{abstract}

\section{INTRODUCTION}\label{sec:intro}
High-precision astrometry with an accuracy in the 0.01-1 milli-arcsecond (mas) range is a powerful technique to detect and characterise extrasolar planets and substellar companions of dwarf stars\cite{Pravdo:2005fu, Casertano:2008th, Sahlmann:2011fk, Lazorenko:2011lr}, especially when complemented with radial velocity measurements. With its sensitivity to the true companions mass, the astrometry technique is particularly suited to constrain the mass distributions of exoplanet populations\cite{Sahlmann2012PhD}. So far, only few instruments are capable of realising the required precision and long-term stability $<1$~mas, but it has been demonstrated e.g. with VLT optical imaging\cite{Lazorenko:2009ph}, the HST\cite{Benedict:2010ph}, and infrared interferometry\cite{Muterspaugh:2010lr}.\\ 
Using the new capabilities offered by the PRIMA facility\cite{Delplancke:2008xr, Belle2008} at the Very Large Telescope Interferometer (VLTI) of the European Southern Observatory, we plan to carry out an astrometric planet search and characterisation programme, which requires an accuracy of 0.01-0.1 mas for the relative position measurement of two stars in a narrow field of view $\lesssim30\arcsec$. We are currently commissioning this mode of operation and once completed, the ESPRI scientific observing programme will begin and the astrometric mode of PRIMA will be offered to the community. PRIMA will then be the unique public facility capable of delivering high-precision relative astrometry to the general user.\\ 
In Section \ref{sec:principle}, we describe the measurement principle and its implementation with PRIMA at the VLTI. Section \ref{sec:obs} discusses the test observations we have carried out on bright visual binary systems and Section \ref{sec:res} presents the results. The status of the science programme of ESPRI is summarised in Section \ref{sec:espri}. We conclude in Section \ref{sec:conclusions} and indicate directions for the immediate future of the project in Section \ref{sec:outlook}.

\section{MEASUREMENT PRINCIPLE AND IMPLEMENTATION}\label{sec:principle}
The high spatial resolution realised by optical interferometers offers the capability of high-precision astrometry. The instantaneous internal delay $w$ that has to be introduced to observe the interference fringe pattern on the detector is related to the star's position $\vec s$ by the relation
 \begin{equation}\label{eq:OPDform2}
 \begin{split}
      w &= \vec{s} \cdot \vec{B}+c_1\\         
          &= \cos E\,\cos A\,\, B_{North}  + \cos E\, \sin A \,\, B_{East} + \sin E\,\, B_{Elev}+c_1
          \end{split}
 \end{equation}
expressed in the horizontal system, where $\vec B = (B_{North}, B_{East}, B_{Elev})$ is the astrometric baseline of the interferometer, $A$ and $E$ are the star's azimuth and elevation, respectively, and $c_1$ is a constant. In this wide-angle mode, the accuracy is limited to a level of $\sim$10 mas\cite{Shao:1990qq} by atmospheric turbulence and by the knowledge of $\vec B$. Both limitations can be overcome by simultaneously observing two stars $\vec s_1$ and $\vec s_2$ having a small angular separation\cite{Shao1992}. In the narrow-angle mode, the atmospheric noise affecting both stars is correlated and mitigated by the differential measurement and the requirement on the knowledge of $\vec B$ is relaxed. The observable becomes 
\begin{equation}\label{eq:dopdmodel}
\begin{split}
      \Delta w  & =  \Delta \vec{s} \cdot \vec{B}+c\\
                  & =  B_\mathrm{North} \, \left( \cos{E_{2}}\, \cos A_{2} - \cos E_{1}\, \cos A_{1}  \right) \\
                  &\, +  B_\mathrm{East}  \, \, \, \left( \cos E_{2}\, \sin A_{2} - \cos E_{1}\, \sin A_{1} \right)   \\
                  &\, +  B_\mathrm{Elev} \, \, \, \left( \sin E_{2}               - \sin E_{1}               \right) +c,\\
                  \end{split}
      \end{equation}
where $\Delta \vec{s}$~is the stars' angular separation vector, $\Delta w$ is the differential delay within the interferometer, and $c$ is a constant. The design goal of PRIMA is to achieve an accuracy of 0.01 mas over a typical field of view of 10 arcseconds with a baseline of $\sim$ 100 m, which translates into a relative precision of $10^{-6}$ or a 100 $\mu$m accuracy requirement on the baseline knowledge and 5 nm accuracy on the differential delay measurement. 
\begin{figure}[h!]
\begin{center} 
\includegraphics[width= 0.7\linewidth]{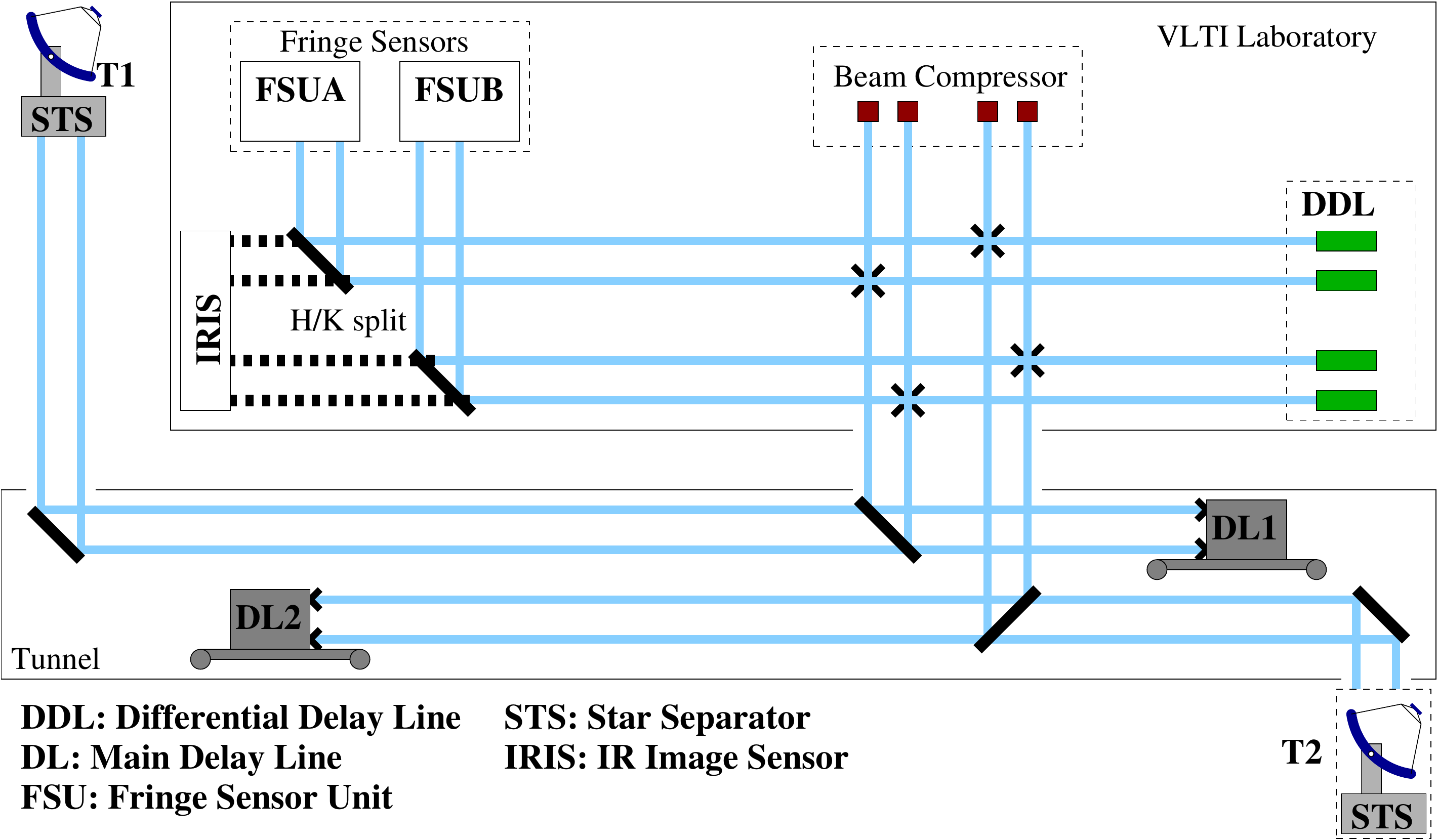} \end{center} 
\caption{Schematic of the stellar beam paths for PRIMA astrometric observations showing the main subsystems. The beams propagate through underground light ducts from the telescopes to the delay line tunnel from where they are relayed into the beam combination laboratory. The metrology monitors the path lengths of the four beams between the respective endpoints in the fringe sensors and the star separators. Figure adapted from [\citenum{Sahlmann2012PhD}].}
\label{fig:dwg}
\end{figure}

It is evident that specialised instrumentation and a detailed calibration plan are necessary to meet these stringent requirements on the instrument stability for an exoplanet orbit measurement programme, i.e. over a timebase of several years. The PRIMA facility realises such an infrastructure and represents a major enhancement of the VLTI at the Paranal Observatory in Chile. It introduced the dual-feed capability and is designed for observation of two stellar objects with an angular separation of $\sim$3-40\arcsec. It consists of star-separator modules at the telescopes (STS\cite{Nijenhuis:2008cy}), differential delay lines (DDL\cite{Pepe2008}), an internal laser metrology\cite{Schuhler2007b}, and the fringe sensor unit (FSU\cite{Sahlmann:2009kx}) in the beam-combination laboratory. The status of the PRIMA facility is presented in [\citenum{Schmid2012}]. Figure~\ref{fig:dwg} shows the schematic beampath of PRIMA operating with two 1.8 m auxiliary telescopes for astrometry: Two stars within the field-of-view are observed and their light is fed into separate beams by the star separator, which splits the image plane between the stars. This module is located below the telescope and includes metrology endpoints. The beams are routed towards the main delay lines, where they experience identical delays, thanks to the dual-feed capability that was anticipated from the beginning of the VLTI project. After pupil diameter reduction to 18~mm, each of the four stellar beams propagates through one differential delay line, thus it is possible to control the delay experienced by each beam individually. Finally the $K$-band light of primary and secondary star is combined in one of the twin detectors of the fringe sensor unit, respectively, which detects the interference signal and drives the fringe-tracking loops\cite{Sahlmann:2009kx}. The $H$-band light is used for slow- and, if applicable, high-bandwidth angle tracking with an infrared camera and image actuators located in the star separator modules. The PRIMA metrology system is launched in the fringe sensor unit and monitors the paths of all four stellar beams simultaneously. The infrared laser beams propagate in the central obscuration of the stellar beams to minimise non-common path errors. Its endpoints are given by the beam combiners in the laboratory and by retroreflectors inside the star separators at the telescopes.

\section{OBSERVATIONS}\label{sec:obs}
\begin{figure}[h]
\begin{center} 
\includegraphics[width= 0.48\linewidth]{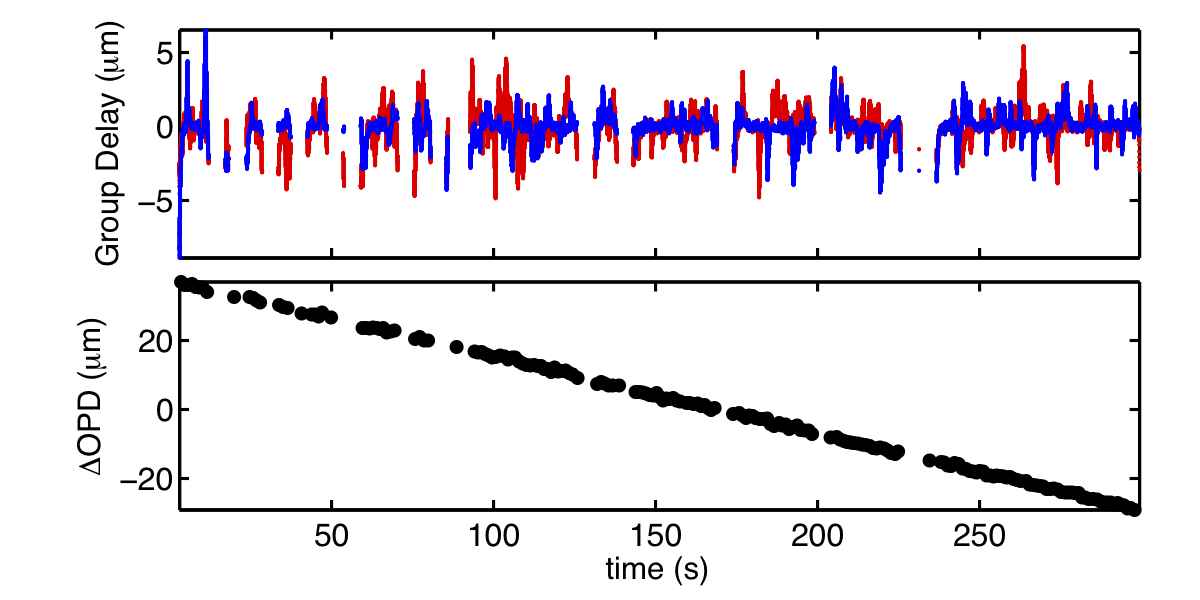} 
\includegraphics[width= 0.48\linewidth]{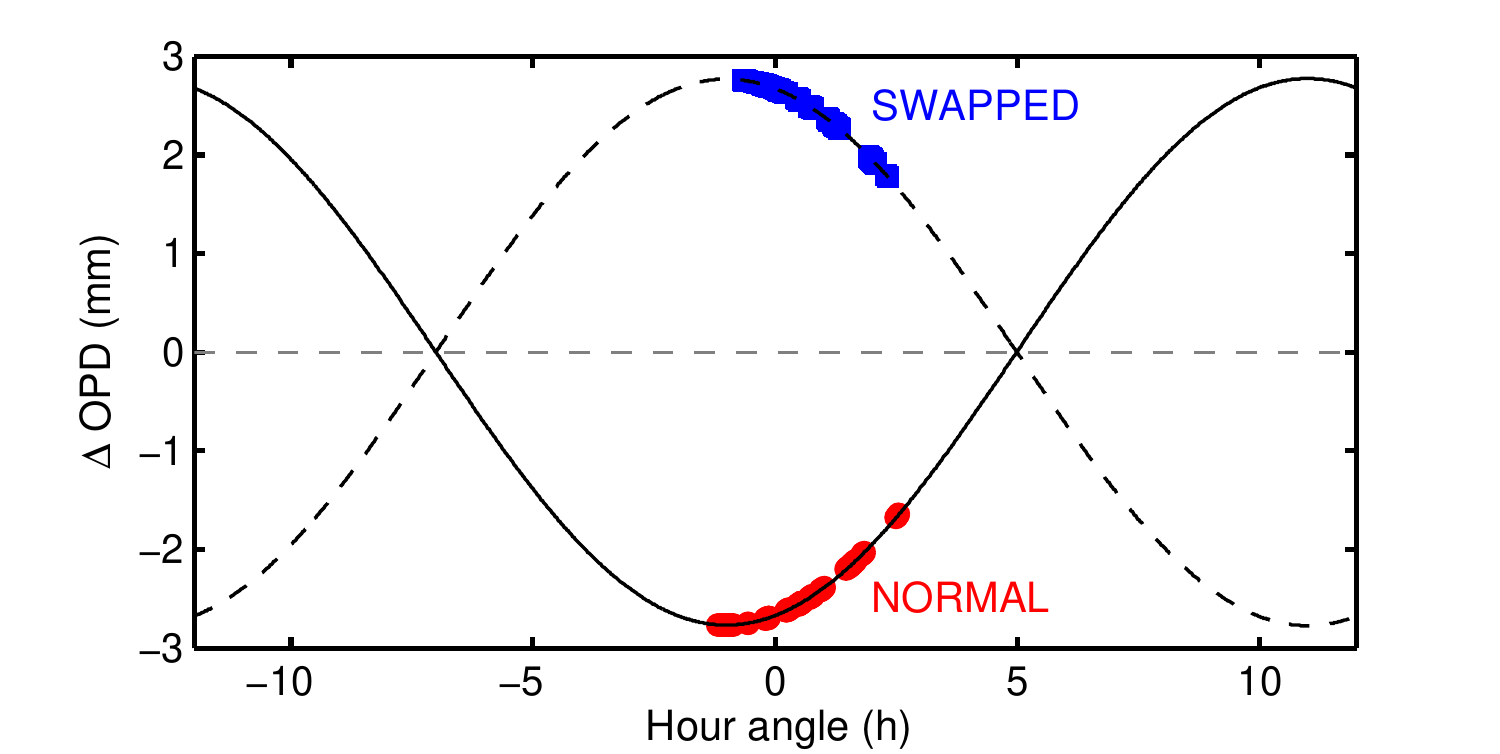}
 \end{center} 
\caption{\emph{Left:} The top panel shows the group delays of FSUB (red) tracking on the primary and FSUA (blue) tracking on the secondary star. The time series interruptions correspond to times where the fringe tracking loop was not closed. The bottom panel shows the change of differential delay ($\Delta \mathrm{OPD}=\Delta w$) as observed with the metrology. The rate of change is of the order 100 nm/s. \emph{Right:} Evolution of differential delay between both interferometer feeds in normal (solid, red) and swapped (dashed, blue) mode. The amplitude is typically in the 0--20 mm range.}
\label{fig:ftk}
\end{figure}

The observation strategy to obtain astrometric measurements with PRIMA consists of simultaneously observing fringes of the primary and the secondary star while the metrology system records the internal differential delay (Fig.~\ref{fig:ftk}). Because the metrology system is incremental, an additional calibration step is required, which consists of 'swapping' the roles of primary and secondary star by rotating the field of view above the field-splitting mirror by 180\degr. This is done with the help of the derotator which is part of the star separator. Consequently, the differential delay between both stars changes its sign and the metrology zero-point can be determined. In addition, any strictly differential error term is mitigated by this procedure. In practice, the array is pointing inbetween the target pair and the slightly off-axis beams are brought back on-axis after field splitting by field mirrors inside the star separators. The stars are acquired by the twin detectors of the fringe sensor unit and the angle tracking camera and after the photometric calibration step\cite{Sahlmann:2009kx}, fringes on the primary star are searched and tracked with one fringe sensor and one main delay line as actuator. Then, fringes on the secondary star are acquired and tracked with the other fringe sensor and one differential delay line as actuator. The passive metrology systems monitors the differential delay in this configuration and data is recorded. The procedure is repeated after the 'swap' of primary and secondary star. An astrometric sequence consists of dual fringe tracking data containing at least one swap operation. The core observable is given by the metrology system, with the consequence that a loss of the metrology signal terminates an astrometric sequence. 
\\
The first astrometric sequence of PRIMA was recorded in January 2011 on the bright visual binary HD\,66598/ HD\,66598\,B ($m_K \simeq 3.0 / 4.6$ mag). Those initial measurements were affected by an error in the metrology software, which introduced large ($>20\,\mu$m) biases in the data. The metrology error was corrected in August 2011 and the earlier data could be corrected retroactively, but large systematic errors persisted on wide-separation targets. During the astrometric commissioning phase, many adjustments to the observing procedure were made to reduce potential error sources (for instance, the derotator motion was restricted to within a minimum range) and to optimise the observing efficiency. Several target pairs over a wide range of separations were observed and we concentrated on acquiring long sequences to observe and track down the systematic errors.

\subsection{Data reduction and analysis}
\begin{figure}[h!]\begin{center}
\includegraphics[width = 0.6\linewidth,  trim = 0cm 0cm 4.5cm 0.8cm, clip=true]{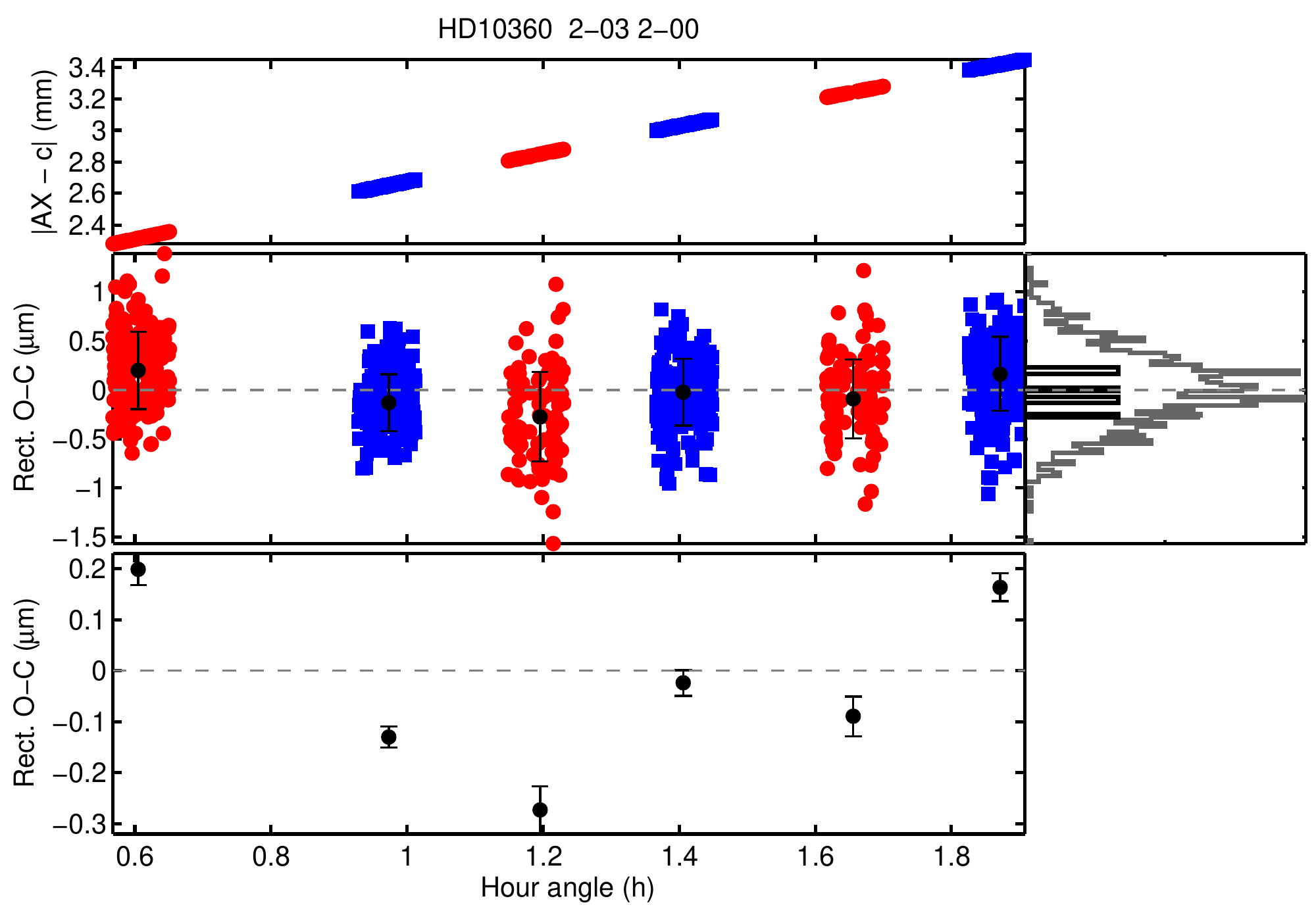}
\caption{PRIMA observation of the small-separation (11\arcsec) target HD\,10360/HD\,10361 on 26.08.2011. \emph{Top}: Absolute value of the measured differential delay $|\Delta w - c|$ as a function of hour angle. Normal mode data is shown in red and swapped mode data in blue. \emph{Middle}: Residuals of the astrometric fit. Each point corresponds to a 1-second average and file-average bins are shown in black. \emph{Bottom}: Binned residuals with error bars that rely on Gaussian statistics.} 
\label{fig:axres1}\end{center}\end{figure}

For the purpose of commissioning and assessing the instrument's performance we use custom data reduction tools and the results shown herein were obtained with the methods described in [\citenum{Sahlmann2012PhD}]. The reduction is based on the real-time group delay computations of the fringe sensors and the estimate of the internal differential delay delivered by the metrology system to obtain the astrometric observable $\Delta w$, downsampled to 1-second averages. The analysis and model fitting is done using the simple model Eq.~\ref{eq:dopdmodel} and assumes that the baseline $\vec B$ is constant in time and identical for wide-angle and narrow-angle observations (the wide-angle baseline was determined from the observation of a set of single stars with accurately known astrometry). We therefore initially neglect any effect related to inaccurate baseline knowledge and to crosstalk between imaging and astrometric baseline. This simplified analysis is suitable for investigating systematic errors that are larger than typically one micron, i.e. exceed the extend of the white fringe in delay space. We note that dispersion and refraction effects on the metrology are negligible to first order, because the differential delay lines are kept under vacuum. 
Figure~\ref{fig:axres1} shows an example of a PRIMA test observation. Typically, we interlace several normal and swapped observations, where the cycle time between both modes is $\sim$15 min and the duty cycle is approximately 50-70\,\%.

\subsection{Quality metrics}
To evaluate the quality of PRIMA astrometric observations, we use three complementary metrics:\begin{enumerate}
  \item Amplitude and structure of the O-C residuals of the astrometry fit using the model Eq.~\ref{eq:dopdmodel} or similar:\\ Large residuals indicate an inappropriate model function and/or excess noise. The residual structure and distribution can give insight into the underlying systematic and allows us to test whether the noise is 'white', i.e. normally distributed. We investigate the residuals of the 1-second averages as well as the binned residuals over one file (typically 5 min of data) to distinguish between measurement/atmospheric noise and systematic errors.
  \item Repeatability of measurements:\\ Under the assumption that the target's separation is not variable, several PRIMA observations of the same target should yield the same separation vector. The anticipated single-measurement precision of 0.01 mas sets tight constraints on the knowledge of the relative motion of the target pair. Physical binaries are suitable targets because the effects of differential parallax and proper motion are small.
  \item Comparison with independent measurements:\\ We observed one target with an independent high-resolution imager (NACO/VLT) to constrain the PRIMA plate scale. Again, the high precision requirement stresses the importance of selecting an appropriate target and observation strategy.
\end{enumerate}

\section{RESULTS}\label{sec:res}
\subsection{Astrometric precision}
\begin{figure}[h!]\begin{center}
\includegraphics[height = 0.36\linewidth,  trim = 1cm 4.4cm 2cm 0cm, clip=true]{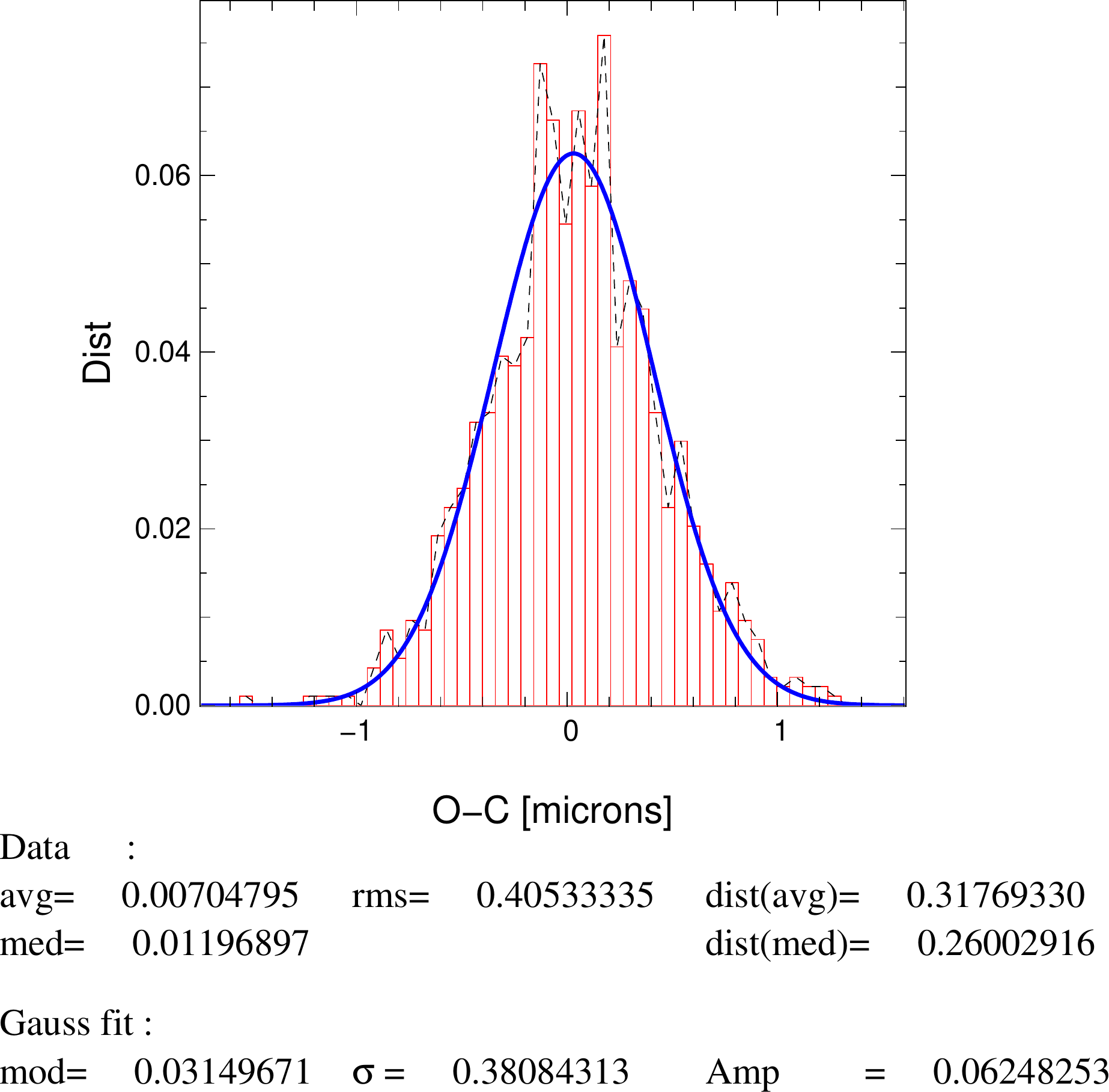}
\includegraphics[height = 0.37\linewidth,  trim = 0cm 0cm 0cm 0cm, clip=true]{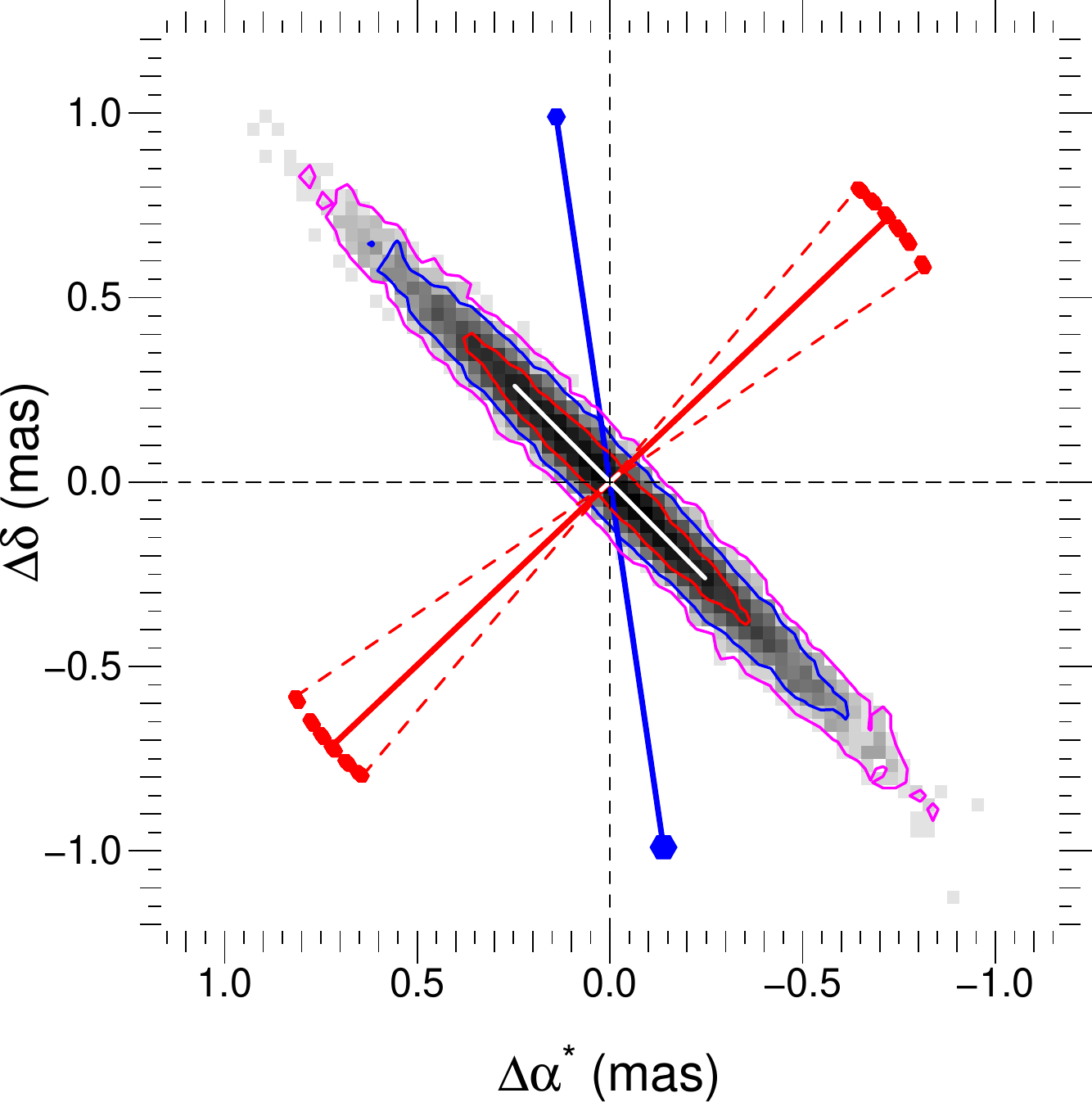}
\caption{\emph{Left:} Residual's histogram of the HD\,10360 observation shown in Fig.~\ref{fig:axres1} and the best Gaussian fit to it. \emph{Right:} The two-dimensional distribution of the relative secondary's position in the sky for 10\,000 Monte Carlo simulations is shown in grey-shading. North is up and East is left. Contour lines indicate 1,2,3-$\sigma$ confidence intervals. The blue line indicates the position angle of the binary, the red circles show the equivalent $u$-$v$-coordinates of the observations, and the solid red line illustrates the average projected baseline orientation. }
\label{fig:precisionHD202730}\end{center}
\end{figure}

For short sequences on small separation targets, the systematic errors are smaller than the measurement noise. For instance, the residual dispersion of the HD\,10360 observation shown in Fig.~\ref{fig:axres1} is 400 nm RMS (unbinned) and their distribution is well approximated by a Gaussian curve (Fig.~\ref{fig:precisionHD202730}). Therefore, we can tentatively assume Gaussian statistics and the total number of 937 one-second data points ($\simeq 0.26$\,h) results in an expected astrometric precision of $\sim$0.030 mas, given the average projected baseline length of 90 m. Using Monte-Carlo resampling, we obtained the distribution of secondary star positions in the sky (Fig.~\ref{fig:precisionHD202730}) and obtain a 1-$\sigma$ confidence interval of 0.032 mas along the average projected baseline orientation.
Applying the theoretical formula of [\citenum{Shao1992}], we obtain the atmospheric limit for the precision of this observation of 0.027 mas. The astrometric precision of PRIMA on bright small-separation binaries is thus only slightly worse than the theoretical expectation. This is a necessary but not sufficient condition for the feasibility of the ESPRI scientific programme, because it is the long term astrometric accuracy that determines the sensitivity to planetary signatures. So far, the PRIMA astrometric accuracy is far worse than the precision, which is discussed in the following sections.

\subsection{Systematic biases}
\begin{figure}[h]
\begin{center} 
\includegraphics[width= 0.7\linewidth]{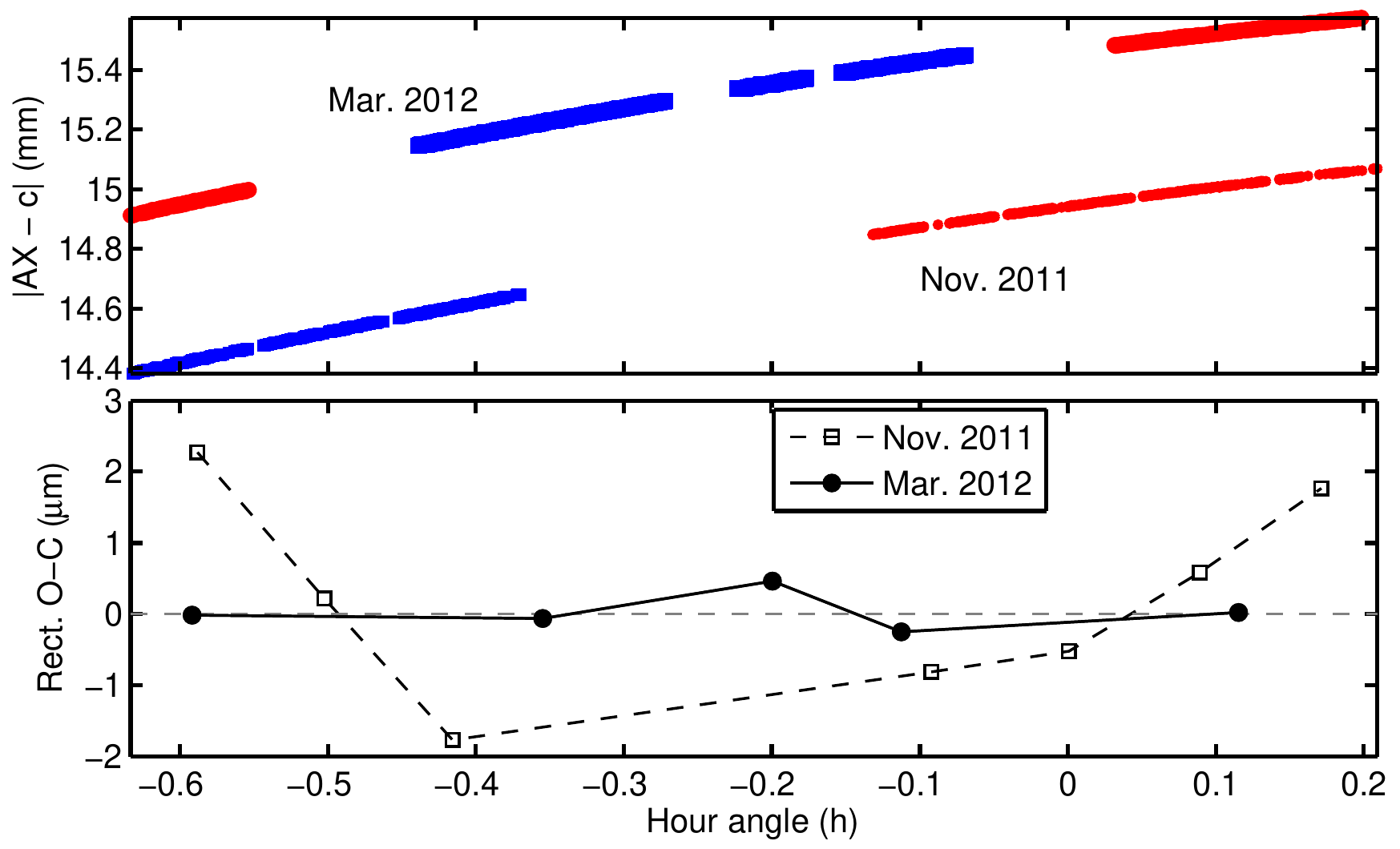} 
 \end{center} 
\caption{Observations of HD\,66598 before and after telescope re-alignment. \emph{Top:} Differential delay of both sequences. For better readability, the Nov. 2011 sequence was shifted by -0.5 mm. \emph{Bottom:} Binned residuals before (dashed line and open squares) and after (solid line and filled circles) the alignment intervention. The binned residual dispersion decreases from 1.46\,$\mu$m to 0.27\,$\mu$m. Error bars are smaller than the symbol size.}
\label{fig:comp}
\end{figure}

For target separations larger than $\sim$10\arcsec~or sequence lengths $\gtrsim2$~h, the systematic errors are larger than the measurement noise. In extreme cases of fast field rotation and a wide-separation target, the residual RMS can exhibit a wavelike pattern with $\sim$10\,$\mu$m RMS. These patterns reveal a correlation with field rotation and show that the introduced biases change most rapidly around the meridian, i.e. when the field rotation is most important. The critical subsystem in this respect is the metrology endpoint at the telescope. In the current implementation (February 2012), the endpoint is located at the level of M9, i.e. any differential delay that is introduced in the stellar beams between the telescope primary mirror and M9 remains unnoticed by the metrology system and will appear as bias in the astrometric measurement. The derotator assembly, a reflective K-shaped prism, is located just above M9 and is therefore not monitored. Furthermore, we discovered important and pointing-dependent obscurations of the stellar pupils and a significant lateral pupil run-out as a function of telescope azimuth, which points at insufficient alignment of the PRIMA/VLTI optics.
A re-alignment of the auxiliary telescopes (primarily of M4) used for PRIMA performed in March 2012 resulted in a significant reduction of residual amplitude. Figure~\ref{fig:comp} shows the effect with the help of observations of HD\,66598/HD\,66598\,B, a 36\arcsec-separation binary, which were selected to lie within the same hour-angle range and with comparable amount of collected data.
Before the intervention, the residuals show a u-shaped pattern with several $\mu$m amplitude typical for observations of a wide-separation binary close to meridian (the pattern extends far beyond the shown hour-angle range, but has been omitted here to permit a proper comparison). After the intervention, the remaining residuals have decreased by more than a factor of five in RMS. This significant decrease of the systematic error supports the idea that it is introduced in the unmonitored part of the optical beamtrain, i.e. above M9. An extension of the metrology endpoint up to M2 or beyond would supposedly mitigate a large fraction of the observed systematic errors.

\subsection{Astrometric accuracy and plate scale verification}
\begin{figure}[h]
\begin{center} 
\includegraphics[width= 0.41\linewidth]{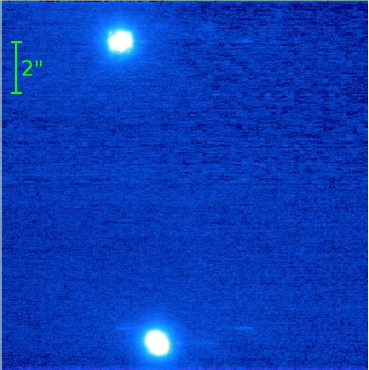} 
\includegraphics[width= 0.5\linewidth]{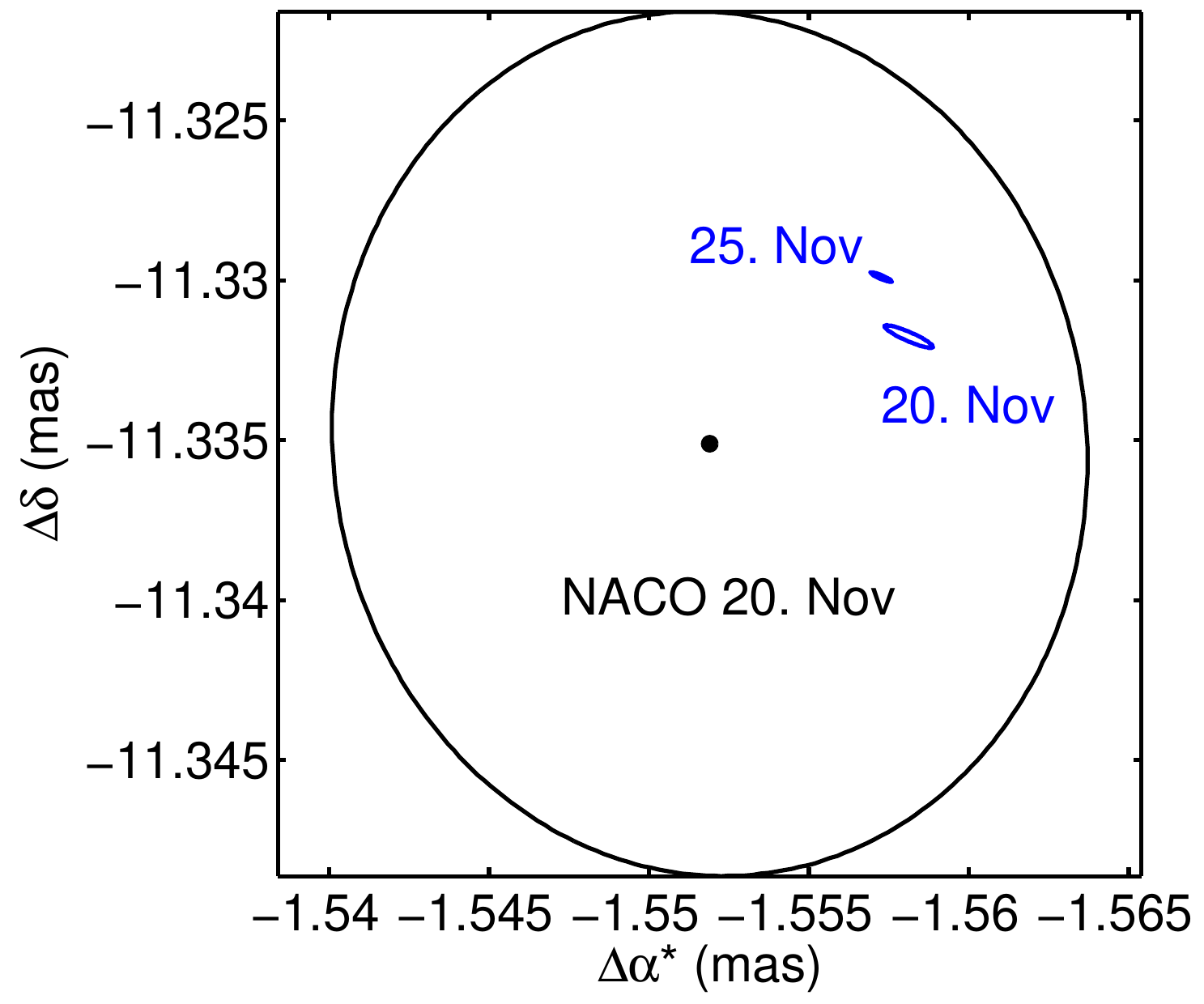} 
 \end{center} 
\caption{\emph{Left:} NACO image of HD\,10360 (top) and HD\,10361 (bottom). \emph{Right:} The separations of HD\,10360/HD\,10361 on 20.11.2011 measured with {\small NACO} (black) and PRIMA (blue) are consistent within the former's error ellipse. The PRIMA measurement obtained in a comparable hour angle range 5 days later on Nov. 25 is also shown in blue.}
\label{fig:NACO}
\end{figure}

To obtain an independent comparison measurement for PRIMA, we observed the HD\,10360 binary simultaneously with PRIMA and the NACO infrared adaptive optics camera of the VLT\cite{Lenzen:2003vn, Rousset:2003ys} on November 20, 2011. We used the 2.17 $\mu$m Br$_\gamma$ filter of the NACO S27 camera to obtain a total of 218 s exposure time in 2000 individual frames. To calibrate the NACO plate scale we subsequently observed an area of the Trapezium cluster and used the astrometry of selected cluster stars published by [\citenum{Close:2012fk}]. Because of the target separation of $\sim$11\arcsec, the plate scale calibration still limits the NACO astrometric accuracy to 13 mas. A sample NACO image and the resulting separation measurement are shown in Fig.~\ref{fig:NACO}.
The simultaneous NACO and PRIMA measurements differ by $\sim$6 mas and are compatible within the error bars, i.e. we can conclude that the PRIMA plate scale is accurate at the 13 mas level on a 11\arcsec~field, i.e. at $10^{-3}$. Since PRIMA is not a field imaging instrument, its \emph{plate scale} refers to the absolute scale fidelity, which for instance can be affected by an error in the astrometric baseline or in the metrology wavelength.\\
With regard to accuracy, the two PRIMA measurements taken 5 days apart differ by $\sim 2$ mas. The expected separation change due to orbital motion of HD\,10360 is $\sim$0.13 mas/day eastward\cite{Sahlmann2012PhD} and can therefore only partly explain the large discrepancy. It is rather attributable to systematic errors of the order of 1 $\mu$m in differential delay, which is the typical amplitude of non-white (and field-orientation dependent) features in the residuals of this target.


\section{THE ESPRI PROJECT}\label{sec:espri}
The primary scientific motivation behind the astrometric mode of PRIMA is to advance our understanding of nearby planetary systems. Once the facility reaches its full potential, an unique region of exoplanet parameter space --- mass, orbital radius, and age --- will be opened for exploration. Hence, the ESPRI project\cite{Launhardt2008} was organised as one large, coherent observational program to exploit PRIMA's narrow-angle astrometry mode for exoplanetary science. 

\subsection{Scientific Goals}
The ESPRI project has identified several goals uniquely enabled by the planned capabilities of PRIMA.
\begin{itemize}
\item Establish the masses of planets already found by radial-velocity surveys. For most of these systems, astrometry is the only feasible way to break the orbital inclination degeneracy. An ensemble of these measurements will refine the planetary mass function, in particular at the upper-mass end where the statistics remain poor. In addition, constraining the orbital inclination of a planet in some specific, enigmatic systems would contribute greatly to physical interpretations.

\item Discover additional, longer-period planets in systems with one or more planets already detected by radial-velocity surveys. An astrometric approach is the most suitable for this task: at a given target distance and planet mass, the sky-projected reflex motion increases with orbital period as $P^{2/3}$, whereas the radial-velocity signal amplitude falls off for wider orbits due to its proportionality to $P^{-1/3}$.

\item Carry out a survey for planets around stars across a broad range of mass and age. Of particular interest are young stars and very nearby main sequence stars of spectral types M through A. Unlike the case for radial-velocity measurements, the high activity of young stars does not corrupt their astrometry (not at the 10 $\mu$as level, at least). Furthermore, the orbital radius range probed by a 5-year astrometry survey complements the inner working angle that limits high-contrast imaging instruments. The most nearby stars---regardless of age---are also of interest to ESPRI, because the astrometric signal is inversely proportional to distance from the Sun.
\end{itemize}

\subsection{Target Selection}
In accordance with the stated goals, the ESPRI programme will target three different groups of stars:
\begin{enumerate}
\item Stars with known (published) exoplanet candidates detected via radial velocity (RV). For these systems, with known periods and predictable minimum astrometric amplitude, we want to resolve the sin\,$i$\ ambiguity and search for additional long-period planets.
\item The most nearby stars within 15 pc of the Sun to survey for planets in the 1-3 AU range and down to the Uranus-mass regime.
\item Young stars (age $<$ 300 Myr) within 100\,pc of the Sun to survey for giant planets (Jupiter-mass) in the 1-3 AU range.
\end{enumerate}
\noindent We began with an initial list of $\sim 1000$\ stars meeting criteria of brightness, Declination, along with published properties matching those appropriate for each category. Next, through a dedicated observing campaign, we investigated the spectroscopic properties and reference star situations of these target candidates. Based on this work, we cut the initial list down to $\sim 260$\ feasible targets. Finally, taking signal-to-noise ratio calculations of expected astrometric signals into account (using the published planet candidate measurements for target group 1, and an assumed set of giant planet orbital parameters for target groups 2 and 3), we further reduced the list to arrive at $\sim 100$\ top ESPRI targets. In particular, we have selected:
\begin{enumerate}
\item 31 stars with published RV exoplanet candidates whose orbits we will have the sensitivity to characterise.
\item 35 nearby main sequence stars for which we will have the sensitivity to detect at least a Saturn-mass planet in a 4-year orbit.
\item 32 nearby young stars for which we will have the sensitivity to detect at least a Jupiter-mass planet in a 4-year orbit.
\end{enumerate}
The actual range of planet parameters that can be explored within the ESPRI program will depend on the yet-to-be determined performance of PRIMA's narrow-angle astrometry mode, and in particular its measurement stability over long time scales. Here we have assumed a simplified error scenario, in which the single-measurement precision has a nominal 1-$\sigma$ value of 20\,$\mu$as at 10 arcsec separation and degrades with increasing reference star separation 
due to anisoplanatism.

\subsection{Observation Strategy}
The ESPRI project will use of order 200 observing nights with the astrometric 2-AT array, spread over 5-8 years, to search for astrometric exoplanet signatures. We currently assume that one full astrometric measurement cycle with 30 min total integration time takes $\sim$1 hr of telescope time, and that on average six such astrometric measurements can be obtained in one observing night. This gives us a total budget of $\sim$1200 astrometric data points. To estimate how many stars we can survey within this total budget, we have performed simulations on certain hypothetical planetary systems. These simulation shows that for an RV-constrained system, we need of order 15 astrometric data points with single-measurement SNR $\sim$10 (along the better-constrained direction, w.r.t. the large axis of the projected orbit) in order to estimate the true inclination with a 1-$\sigma$ relative uncertainty of better than 20\,\%. Hence, we will need a total of about 100 nights in order to astrometrically characterise $\sim$30 RV targets. For targets that have no RV constraints, the possible presence of a planet and its orbital period are a priori unknown. We assume that we need of order 15 data points per star, distributed over the first year of observations, in order to sufficiently constrain the differential parallax and search for residual signals. In the case of significant residuals that hint towards a planet, we need to schedule at least 15 more measurements distributed well over the predicted orbital period in order to derive an orbital solution. We therefore estimate that we can survey about 30 more stars from our two non-RV lists (nearby and young).\\
The first scientific results from early ESPRI observations can only be expected on a timescale of 1-2 years from the start of astrometric measurements for two reasons: (i) planets (both known and yet unknown) with orbital periods significantly shorter than 1 yr produce astrometric signals with amplitudes significantly smaller than 50-80 muas, and (ii) we have to solve for the differential parallax with the reference star (except for the few cases when the two stars are physically bound).

\subsection{Data reduction pipeline}
The ESPRI consortium is also in charge of providing the Astrometric Data Reduction Software (ADRS), which reduces observations obtained with PRIMA. It will be integrated into the existing ESO software infrastructure and thus 
conforms to ESO standards.\\
The ADRS consists of two parts, the online pipeline which applies initial data selection and averaging, and the offline part which applies numerous instrumental, environmental and astrometric corrections. The instrumental corrections include corrections for the relative phases between the four quadrants of the FSU, dispersion corrections of the differential and main delays measured with PRIMET, and the removal of the zero-point with the help of observations obtained in normal and swapped mode, respectively. Astrometric corrections include corrections for proper motion, annual and diurnal parallax, perspective acceleration, relativistic aberration, light time delay, relativistic light deflection by solar system bodies, and for Earth orientation in general.\\ 
The main purpose of the pipeline is the reduction of observations obtained in narrow-angle mode, i.e. with dual-star interferometry, but it can also be used to obtain the astrometric baseline from observations of single stars. At present, the pipeline is functional, but not all algorithms are final yet. They will have to be adapted to the way PRIMA is operated once that is finalised. 

\section{CONCLUSIONS}\label{sec:conclusions}
\begin{enumerate}
  \item The PRIMA dual-feed facility has successfully been integrated in the VLTI and is operational for astrometric observations. Several technical runs and three astrometric commissionings have taken place in 2011/12 with the goal of establishing the instrument's astrometric capabilities.
  \item The astrometric precision of PRIMA lies in the expected range and a performance of 0.03 mas on bright small-separation ($\lesssim$10\arcsec) binaries has been achieved. 
  \item For wide-separation binaries ($\gtrsim$10\arcsec) and long ($>$2 h) observing sequences, large systematic errors that are correlated with the field rotation are observed. Those errors can amount to several tens of micro-meters corresponding to tens of milli-arcseconds in astrometry.
  \item By comparison with NACO observations, we find that the plate scale of PRIMA is accurate at better than $10^{-3}$. 
  \item The astrometric accuracy of PRIMA in the current (February 2012) state is limited to a few milli-arcseconds by systematic errors. There is evidence that those errors originate in the telescope beam train between the primary mirror and the M9 mirror, which is not monitored by the laser metrology system. The decision to place the metrology endpoint at M9 was taken during the design phase of PRIMA to limit the impact on the operational infrastructure of the VLTI. A design with the endpoints at the secondary mirror was proposed at the time, but deferred until proven necessary to reach the required astrometric accuracy\cite{Delplancke2006}. A hardware modification to place the endpoints at the level of the secondary mirror is currently being investigated.
  \item The astrometric performance of PRIMA as of February 2012 is insufficient to allow us to begin with the ESPRI programme for extrasolar planet search and characterisation.
  \item The preparation of the ESPRI target list has been concluded and we will begin with scientific observations of exoplanet targets when the long-term astrometric accuracy of PRIMA will have reached the $\sim$0.1 mas level on targets with $m_K=7-8$ and $\Delta m_K=2-6$.
  \end{enumerate}

\section{OUTLOOK}\label{sec:outlook}
The immediate future of the PRIMA and ESPRI projects will be focussed on establishing the astrometric performance necessary for the ESPRI science programme. The dominant systematic errors are supposedly mitigated by improving the alignment of the PRIMA and VLTI opto-mechanics and by moving the metrology endpoints closer to the telescope entrance pupil. We expect that these actions will result in a reduction of the systematic errors into the sub-micron regime. At this level of precision many more effects will come into play and will have to be considered carefully. At the fringe detection, biases introduced by changes in the photometric and spectral response of the facility have to be considered and calibrated. The influence of differential atmospheric dispersion in the air-filled VLTI beam-trains on the astrometric measurements have to be accounted for. The determination of the astrometric narrow-angle baseline with a relative accuracy of $10^{-5}-10^{-6}$ will require detailed modelling of the telescope and the relative position of the metrology endpoints. In addition the baseline calibration plan has to be refined. All of these efforts are underway and the demonstrated progress in the systematic testing of the facility is promising. In parallel to the improvement in astrometric accuracy, the operational limits in terms of limiting magnitude, admissible magnitude difference between the two targets, and the performance dependency on ambient conditions have to established. The operation of the PRIMA facility has proven to be robust, but its efficiency can further be improved by optimising the observing procedure and the fringe tracking control system. As demonstrated by the successful identification of the factors limiting the PRIMA performance so far, the putting into operation of a multi-layered facility like PRIMA/VLTI relies on intensive interdisciplinary collaboration as realised between the partners of the ESPRI consortium and ESO in Europe and Chile within the Data Analysis Working Group (ESPRI-DAWG).

\acknowledgments     
The authors would like to express their gratitude to the technicians, engineers, and scientists of the PRIMA, VLT/VLTI, and ESPRI teams that made the deployment of the PRIMA facility possible. We thank M. Accardo, J. Alonso, L. {Andolfato}, H. {Baumeister}, P. {Bizenberger}, P. Bourget, A. {Cortes}, T.~P. {Duc}, F. {D\'erie}, N. {Di Lieto}, M. {Fleury}, R. {Frahm}, B. {Gilli}, P. Gitton, N. Gomes, S. Guisard, P. Haguenauer, L. Jocou, A. Jost, S. {L\'ev\^eque}, C. {Maire}, D. {M\'egevand}, S. {M\'enardi}, A. M\"uller, S. {Morel}, J. Ott, R. {Palsa},  E. Pedretti, I. {Percheron}, A. {Pino}, D. Popovic, E. Pozna, F. Puech, A. Ramirez, F. Somboli, I. {Stilz}, G. {Valdes}, G. Van Belle, L. Weber, and many others. We thank A. Kaufer for granting us NACO technical time and D. Mawet and L. Tacconi-Garman for kindly preparing and executing the NACO observations.


\bibliography{SPIE2012}   
\bibliographystyle{spiebib}   

\end{document}